# Using directed acyclic graphs to determine whether multiple imputation or subsample multiple imputation estimates of an exposure-outcome association are unbiased

**Authors**: Paul Madley-Dowd[1,2,3,4], Rachael A. Hughes[1,2], Maya B. Mathur[5], Jon Heron[1,2,4]*, Kate Tilling[1,2]*

\* Joint Senior Authors

**Affiliations**:

[1] MRC Integrative Epidemiology Unit at the University of Bristol, United Kingdom

[2] Population Health Sciences, Bristol Medical School, University of Bristol, United Kingdom

[3] NIHR Biomedical Research Centre, University of Bristol, Bristol, United Kingdom

[4] Centre for Academic Mental Health, Population Health Sciences, Bristol Medical School, University of Bristol, United Kingdom

[5] Quantitative Sciences Unit, Department of Medicine, Stanford University

# Abstract

Missing data is a pervasive problem in epidemiology, with multiple imputation (MI) a commonly used analysis method. MI is valid when data are missing at random (MAR). However, definitions of MAR with multiple incomplete variables are not easily interpretable and graphical model-based conditions are not accessible to applied researchers. Previous literature shows that MI may be valid in subsamples, even if not in the full dataset. Practical guidance on applying MI with multiple incomplete variables is lacking. We present an algorithm using directed acyclic graphs to determine when MI will estimate an exposure-outcome coefficient without bias. We extend the algorithm to assess whether MI in a subsample of the data, in which some variables are complete, and the remaining are imputed, will be valid and unbiased for the same coefficient. We apply the algorithm to several simple exemplars, and in a more complex real-life example highlight that only subsample MI of the outcome would be valid. Our algorithm provides researchers with the tools to decide whether (and how) to use MI in practice when there are multiple incomplete variables. Further work could focus on the likely size and direction of biases, and the impact of different missing data patterns.

**Keywords**: Statistical methodology; Missing Data; Selection bias; Directed Acyclic Graphs

# Background

Missing data is a common problem in epidemiological studies. In this paper, we focus on the case where the target parameter is the effect of an exposure on an outcome conditional on a set of covariates, and some of the analysis model variables are incomplete.

The simplest approach to analysing incomplete data is to use complete records analysis (CRA) where all study units with missing data in any variable are excluded. The estimated target parameter is unbiased using CRA when the probability of inclusion in the analysis is independent of the outcome variable conditional on the covariates [1,2]. Directed acyclic graphs (DAGs) [3,4] can be extended by including a missingness indicator for each incomplete variable (also called m-DAGs) [5,6], and used to determine whether a target parameter is both recoverable [5,7,8] and estimated without bias by CRA [2]. However, even where the estimate is unbiased using CRA, there may be loss of efficiency compared to the full-data estimate (i.e. if there had been no missing data), due to the reduced sample size [9].

Multiple imputation (MI) involves creating multiple datasets in which every unobserved value is replaced by an imputed (predicted) value, fitting the analysis model in each imputed dataset and combining these parameter estimates using Rubin's rules [10]. Imputation models must include all the analysis model variables and may also include auxiliary variables. Multiple imputation may improve efficiency, by using the partial information available from individuals with incomplete data, and may also estimate the target parameter without bias in situations where estimation using CRA is biased [11]. The target parameter will be estimated without bias using MI if data are missing at random (MAR), conditional on the observed values of the variables included in the imputation model – i.e. the probability of the realized missingness pattern does not depend on unobserved data, conditional on the observed data [10,12]. MI is thus valid, i.e. guaranteed to yield unbiased estimates of the conditional expectation of $Y$ given $X$ and $W$ for any correctly-specified analysis model, if data are MAR. Guidelines for dealing with missing data, or for implementing MI, tend to emphasise the need to assess the plausibility of the MAR assumption – but give no details on how this might be done [11,13-17].

Explanation of the MAR assumption is usually given in context of a single incomplete variable, which is easily misunderstood or misapplied when there are multiple partially observed variables resulting in different missing data patterns. For example, consider two partially observed variables $X$ and $Y$, where missingness in $X$ does not depend on any other variable and missingness in $Y$ depends on $X$ (Figure 1A, for expanded discussion see Supplementary). Among individuals with an observed $X$, missingness in $Y$ is independent of unobserved data given observed data ($X$). However, among individuals with a missing $X$, missingness in $Y$ depends on unobserved data ($X$), so the example in Figure 1A would be described as missing not at random (MNAR; i.e., not MAR or missing complete at random, MCAR).

Part of the difficulty in assessing MAR even in such simple examples lies in the definition originally given by Rubin [12] and its original intended purpose [18]. The shortcomings of the definition have been explored in detail elsewhere, with several alternative definitions proposed, including "everywhere-MAR" [18-23]. Additionally, MAR as expressed in terms of the distribution of unobserved data cannot be assessed using only the observed data. Graphical approaches, based on DAGs [3] [4] have been used to define MAR (v-MAR) and related concepts, and provide conditions for valid MI [8]. Mathur and Shpitser have provided a "z-MAR" definition, which holds graphically when the "m-backdoor criterion" also holds, and allows for valid imputation [23]. In words, "the m-backdoor criterion holds if any paths between missingness indicators and incomplete analysis variables are blocked conditional on both the modelled [and complete] auxiliary variables and the complete analysis variables" [23]. The graphical definitions of MAR (z-MAR, v-MAR) are related to Rubin's definition, in that v-MAR implies Rubin-MAR [21], and v-MAR is a special case of z-MAR with no auxiliary variables [23]. However, none of these graphical approaches are described in a way that is accessible to applied researchers.

As well as little information on how to assess plausibility of MAR in practice, there is no guidance on how to proceed if MAR is not considered plausible, other than to use sensitivity analyses which are mostly framed in terms of "MNAR" as specifying differences in distributions between different missing data patterns. Little and Zhang have developed a "subsample MI" approach, whereby the estimated target parameter may be unbiased using MI restricted to individuals with observed values for some variables, even when it is biased using MI of the whole dataset [24]. They show that the conditional exposure-outcome association (often estimated as a regression coefficient), is unbiased using subsample MI when:

1) Within the subsample, the data are MAR (the authors used Rubin-MAR).
2) The probability of inclusion in the subsample does not depend on the outcome variable conditional on the covariates in the analysis model (including the exposure).

Intuitively, the first condition ensures that MI is valid within the subsample, and the second that the subsample estimate of the target parameter is an unbiased estimate of the full data parameter. The first condition would apply also to other target parameters (e.g. the mean of the outcome), but the second condition is specific to estimation of an exposure-outcome association [25].

The purpose of the current paper is to bridge the divide between applied researchers and methodological foundations for validity of MI, by providing an algorithm, and showing how to operationalise it using DAGs, to 1) assess whether MI applied to the whole dataset can estimate the exposure-outcome association without bias and, if not, 2) identify whether the association is estimated without bias using MI applied to a subsample of the data.

# Motivating example

We demonstrate the issues facing the applied researcher with a previously applied example [25] from the Avon Longitudinal Study of Parents and Children [26-28]. The analysis model was a linear regression of offspring intelligence quotient (IQ) scores at age 15 (the outcome) on maternal smoking during pregnancy (the exposure), adjusted for maternal age, education and parity during pregnancy (all proxies proxy for socioeconomic position; SEP) [29-33], and offspring sex. The first three adjusted variables are probable confounders; offspring sex is not a confounder but may explain some of the variation in IQ [34]. Linked education score at age 16 was available as an auxiliary variable for IQ at age 15. The exposure, outcome, SEP confounders and auxiliary variable were all incomplete, while offspring sex was complete.

Figure 2 shows a (simplified) DAG of the assumed relationship between variables and missing data indicators. On the basis of prior research [35] we include arrows from SEP confounders to all response indicators, from maternal smoking during pregnancy to its own response indicator and to that for offspring IQ, and from offspring sex to the response indicator for offspring IQ. Following the TARMOS framework [14], we would first conclude that a CRA was likely to be unbiased, as the analysis outcome is independent of the response indicators, conditional on the exposure and covariates. Assessing whether the data might plausibly be MAR (by some definition) given the DAG requires assessing paths between incomplete variables and response indicators, conditional on complete variables. Applying the m-backdoor criterion to the DAG [23], we can see that the data are not z-MAR as there are open paths between incomplete variables and missingness indicators that are not blocked by conditioning on complete analysis model variables or complete auxiliary variables. MI in this example will therefore not be valid. However, applied researchers not familiar with the graphical models literature (i.e. m-DAGs) may find it challenging to arrive at this conclusion. While sensitivity analyses have been suggested if the MAR assumption does not hold [14], guidance so far has not incorporated the exploration of whether the z-MAR assumption may hold in a subsample.

# Establishing the validity of multiple imputation and subsample multiple imputation

In this paper, the target parameter is the association of $Y$ on $X$ conditional on variables $W$. Auxiliary variables $A$ (which are not in the analysis model) may be included in the imputation model. Any of these variables {A, W, X, Y} may be incomplete, with response indicator $R_J$ for variable $J$, where $R_{Ji} = 1$ if variable $J$ is observed for individual $i$, and 0 otherwise, and $R$ the set of response indicators. We define any variables $V_1$ and $V_2$ as *dependent* conditional on $V_3$ if the underlying variables $V_1$ and $V_2$ (not the observed, potentially incomplete, variables) are d-connected, conditional on $V_3$ – i.e. "$V_1$ and $V_2$ are dependent" does not mandate a direct causal path between $V_1$ and $V_2$ [3,4]. We assume that the researcher chooses auxiliary variables *a priori* and includes the same set of variables throughout.

Our primary aim is to establish whether MI applied to all observations in the dataset (i.e. imputing all incomplete variables) can estimate the target parameter without bias. We assume throughout that the MI is carried out correctly, including that all imputation models are correctly specified, and imputation and analysis models are compatible [16]. Our proposed algorithm will suffice for MI to be unbiased for exposure-outcome associations estimated using any type of analysis model, although there are additional scenarios (e.g., outcome-dependent sampling) in which MI will still be unbiased when using a specific analysis model (e.g., logistic regression).

The logic of the proposed algorithm is that MI will identify the joint distribution of $(Y, X, W)$ if all incomplete variables are independent of all response indicators $R$, conditional on complete variables [10]. In the terminology of causal graphs, the MI estimate will be unbiased if the complete variables d-separate all incomplete variables from all response indicators [23].

DAGs provide a useful way to identify sufficient sets of subsampling variables. We use some extensions to usual DAG conventions to allow complete/incomplete variables to be easily distinguished:

> D0) draw the DAG showing assumed causal relationships between all variables to be included in the imputation model(s). Include all analysis model variables $\{X, Y, W\}$, auxiliary variables $A$, and the set of response indicators $R$ for each incomplete variable in $\{X, Y, W, A\}$ [5]. Include unobserved variables $U$ as appropriate (e.g. unmeasured common causes of variables in the analysis model and response indicators) [36]. We let $Z$ be the set of complete variables, and $Z'$ the set of incomplete variables. Identify which variables are complete variables $Z$ (e.g. using a red box), and which are incomplete variables $Z'$ (e.g. using a green outline). Note that unmeasured variables $U$ are considered incomplete (i.e. included in $Z'$), but do not have a response indicator, can never be conditioned on, and will never be imputed.

D1) Identify all variables with an open path to *any* response indicator in $R$, conditional on complete variables $Z$ (i.e. conditioning only on those variables with a red box). This is the initial set $\Phi$. Identify this set, using for example a blue outline for each member of $\Phi$. If $\Phi$ is empty (i.e., no blue outlines) then move on to step D3, otherwise go to next step, D2.

D2) Remove from set $\Phi$ all complete and all unmeasured variables (i.e. $\Phi$ now consists only of those measured variables with both a blue and a green outline, and not any variables with a red box). $\Phi$ is now the set of MNAR-inducing measured variables.

D3) If $\Phi$ is empty (i.e. no variables have a blue and a green outline), then the m-backdoor criterion is met, the data are z-MAR, and the MI estimate using the whole dataset will be unbiased.

If the algorithm above indicates that MI applied to the full data set may be biased, the second stage is a more general procedure to identify whether we could partition the measured incomplete variables into subsets $P$ and $Q$, such that imputing variables $P$ within the subsample of individuals with complete data on $Q$ will be unbiased. The subset $Q$ must contain only variables with a response indicator that is independent of $Y$ conditional on analysis model variables $X$ and $W$. Unmeasured variables $U$ are excluded from both $P$ and $Q$. The process for any given choice of $Q$ is as follows:

D4) Remove all blue and green outlines from all variables in the DAG, keeping the red outlines. Next add red outlines to all variables in $Q$ (as these are "complete" in the subsample as defined above). Add green outlines to all variables in $P$. Condition on all response indicators for variables in $Q$ (draw boxes around them to indicate this, as usual convention. We recommend also indicating that these response indicators are set to 1 in this subsample).

D5) Identify any open paths between variables in $P$ and response indicators for variables in $P$, conditional on $Z$ and $Q$ – i.e. apply steps D1-D2 to identify any open paths between incomplete variables and response indicators that are not blocked by variables with a red box.

D6) If there are no such open paths, then MI of the variables in $P$, applied to the subset of individuals with complete $Q$, will be unbiased.

D7) If there are such open paths, then MI of the variables in $P$, applied to the subset of individuals with complete $Q$, may be biased. Alternative approaches could include considering different subsamples or identifying further auxiliary variables to block paths between incomplete variables and response indicators.

A non-graphical version of this algorithm is also available in the Supplementary Material to aid analysts unfamiliar with DAGs.

## Worked examples of implementation of the algorithm

We apply the algorithm to three scenarios (Figures 3A, 3B and 4), with an accompanying small simulation study described in the Supplementary Material.

In scenario 1 (Figure 3A), $X$ causes $Y$ and $R_Y$, there are no causes of $R_X$. In steps D0-D3 we identify that $\Phi$ consists of variables $X$ and $Y$ and therefore that MI will not be valid using all available data to impute Y and X (because $\Phi$ is not the empty set). We next examine whether there are any subsamples in which MI would estimate the target parameter without bias.

Because MI relies on z-MAR, the aim is to restrict attention to subsamples in which no incomplete variable is related to response indicators. We could investigate imputing $Y$ in the subsample in which $X$ is complete, or vice-versa. $Y$ is not associated with the response indicator for either $X$ or $Y$, conditional on $X$, so either choice is admissible. Following steps D4-D6 indicates that MI restricted to a subsample of either complete $X$ ($Q = \{X\}$ and $P = \{Y\}$) or complete $Y$ ($Q = \{Y\}, P = \{X\}$) will be unbiased. Thus, in this example, we have a choice of strategies. The analyst could carry out both options and compare the results, or might favour the option with the largest sample size. It should be noted that imputing only the outcome without any auxiliary information will not be more statistically efficient than CRA [2] so in this example we have preference for using $Q = \{Y\}$ and $P = \{X\}$.

In scenario 2 (Figure 3B). $X$ causes $Y$, $Y$ causes $R_X$ and there are no causes of $R_Y$. This figure follows similar logic to that above, with the conclusion that the target coefficient can be estimated without bias using MI of $X$ applied to the subsample in which $Y$ is fully observed. However, the estimate from MI of $Y$ applied to the subsample in which $X$ is fully observed may be biased because the response indicator for $X$ is dependent on $Y$.

In scenario 3, $W$ is a confounder of $X$ and $Y$, with $W$ causing missingness in $X$, $X$ causing missingness in $W$, and missingness in $Y$ is independent of all other variables. Figures 4A-4C show the implementation of the algorithm. In steps D0-D3 we identify that $\Phi$ contains variables $W$, $X$ and $Y$, and therefore that MI using all available data is not valid (because $\Phi$ is not the empty set).

We now need to decide how to separate the incomplete variables $Z'$ into sets $P$ and $Q$. As no response indicators are dependent on the outcome given analysis model covariates $X$ and $W$, we can explore multiple possible sets $P$ and $Q$. We could choose variable $X$ as the "complete" variable in step D4 (Figure 4B), i.e. partition incomplete variables into $P$ and $Q$ where $Q$ is "complete" ($X$) and $P$ is "to be imputed" ($Y$ and $W$). Following step D6 we identify that there is no open path from $Y$ or $W$ to $R_Y$ or $R_W$ and thus the subsample MI estimate where $Y$ and $W$ are imputed in the subsample with complete $X$ is unbiased.

Similar logic shows that if we instead choose $W$ as Q in step D4 (Figure 4C) then imputation of $Y$ and $X$ in the subsample with $W$ complete will estimate the target parameter without bias. MI in the subsample within which Y is complete is not valid as the data are not z-MAR within the subsample (there are open paths between incomplete variables $W$ and $X$ and response indicators). It is also possible to unbiasedly estimate the target parameter using both $X$ and $W$ as "complete" variables and imputing $Y$, though this would be less statistically efficient than choosing either $X$ or $W$ as the complete variable.

## Further examples

Figure 5 shows the DAGs for further examples, which are discussed in more depth in the Supplementary Material (alongside a brief simulation study and an application of the algorithm to the canonical DAGs presented by Moreno-Betancur et al.[5 6]). Figure 55A-5C use the same analysis model variables as the example presented in Figure 4, but different missingness mechanisms. Briefly, application of our algorithm suggests the following for each example:

In Figure 5A, $X$ is complete, and $Y$ and $W$ are incomplete with $W$ causing $R_Y$, and $Y$ causing $R_W$. MI applied to the whole sample may be biased. The target parameter is estimated without bias using MI restricted to the subsample where $Y$ is complete, but not using MI restricted to the subsample where $W$ is complete. This is an example where MI in the subsample in which the outcome is complete is unbiased.

Figure 55B is the same as Figure 5A, except instead of a direct path from $Y$ to $R_W$, there is an unmeasured common cause $U$ of $Y$ and $R_W$. As with Figure 5A, MI applied to the whole sample may be biased. However, MI in the subsample where $Y$ is complete may also be biased, because when $Y$ is complete, this opens a path from $W$ to $R_W$ conditional on all complete variables ($X$ and $Y$).

In Figure 5C $X$, $Y$ and $W$ are incomplete, with $R_X$ and $R_Y$ caused by $W$ and $R_W$ caused by $Y$. MI in the full sample is not valid. Applying MI to impute $W$ in the subsample with complete $Y$ and $X$ (i.e. restricting to those individuals with both $Y$ and $X$ fully observed), can unbiasedly estimate the target parameter. This scenario shows that it is sometimes necessary to subsample on observed values of multiple incomplete variables to obtain an unbiased estimate via subsample MI.

## Application to the motivating example

We now implement the algorithm using the applied example described earlier (see DAG presented in Figure 2 and Supplementary Material for figures showing our implementation of the algorithm to this DAG). Following steps D0-D3 we identify the set $\Phi$ = {maternal smoking during pregnancy, IQ at age 15, SEP confounders, linked education score} (i.e. the set of MNAR-inducing variables). As $\Phi$ is not the empty set, MI applied to the whole dataset is not valid. In steps D4-7 we set $Q$ = {maternal smoking during pregnancy, SEP confounders} and $P$ = {IQ at age 15, linked education score}. Here we note that the outcome variable is d-separated from all response indicators for variables in $Q$ conditional on complete variables and variables in $Q$. Further, we cannot use $Q$ = {SEP confounders} as there is still an open path from maternal smoking during pregnancy to its own response indicator. As there are no open paths from the set $P$ to response indicators for variables in $P$, MI of IQ at age 15 and linked education score in the sample with complete SEP confounders and maternal smoking during pregnancy will be valid.

As explored in our previous study [25], we believe it is likely that the response indicators for IQ at age 15 and linked education score are caused by IQ at age 15 and linked education score respectively. Applying the algorithm to the DAG incorporating these relationships would indicate that neither MI applied to the full dataset, nor MI applied to any subsample, would be valid.

# Discussion

We have built on previous work in DAGs [2 5 6], z-MAR and the m-backdoor criterion [23], and subsample MI [24], to provide guidance for deciding how (or whether) to apply multiple imputation to estimate an exposure-outcome association. This fills a crucial gap in the advice to applied researchers on how to deal with missing data, and how to implement MI [14]. The key point is that MI is valid if all incomplete variables are independent of all response indicators, conditional on the complete variables. Identifying a subsample within which all incomplete variables are independent of response indicators, given complete variables, can lead to unbiased estimates of the exposure-outcome association, even where neither CRA, nor MI on the full sample, would do so.

If $Y$ cannot be d-separated from its own missingness indicator by the complete variables – for example, because $Y$ is the direct cause of its own missingness, or if there is an unmeasured common cause of $Y$ and missingness in $Y$ – then typically both CRA and MI (in any subsample of the data) will yield biased estimates of the regression coefficient for the effect of $X$ on $Y$. There are well-known exceptions for certain models, though, such as logistic regression. Thus, one preliminary step before commencing the algorithm could be to examine the DAG and assess whether $Y$ is d-connected to its missingness indicator either directly or via an unmeasured common cause. However, some other scenarios where $Y$ is d-connected to its own missingness indicator could lead to unbiased estimates in subsamples - for example, if incomplete $X$ causes missingness in $Y$ (as in Figure 1A and 3A).

This algorithm will be most useful when there are multiple variables with missing data, and arbitrary missing data patterns. This commonly occurs in the analysis of data from cohort studies, especially where data from different waves of data collection are used. Often it is plausible that baseline covariates (such as measures of economic hardship) may affect likelihood of responding at any wave [35]. If these measures are themselves incomplete (so the data potentially not z-MAR), then MI restricted to those with complete data on economic hardship may be unbiased. Linkage of cohorts to other, more complete sources (e.g. electronic healthcare records) means that restricting MI to samples with complete outcome data may be useful [37]. In RCTs, baseline covariates are usually complete, and the concern is about missing outcome data. Intermediate measures of the outcome may be used to impute the final outcome – either overall, or separately within each arm of the trial [38]. If these variables are also incomplete, then again, our algorithm may be useful to decide on the best analysis strategy.

As with any other methodology based on DAGs (such as confounder selection, or assessment of plausibility of bias), the conclusions depend critically on the DAG assumed. These assumptions should be justified based on external knowledge or prior information (such as documented reasons for missingness). Temporality may help to eliminate some paths – for example, if $Y$ is measured at 12-month follow-up then there cannot be a direct arrow from $Y$ to the response indicator for a variable measured at baseline. Tests of whether the DAGs are incompatible with the dataset can be used [39], and sensitivity analyses conducted to examine robustness to assumptions about which there is uncertainty [40]. As recommended by Moreno-Betancur et al, conclusions should be examined across a range of plausible m-DAGs [5,6].

There are some limitations to our work. MI is not the only method for dealing with incomplete data – others include inverse probability weighting and maximum likelihood methods, which we have not examined here [41]. For MI to unbiasedly estimate the target parameter depends on more than just the conditions outlined here – we have assumed throughout that all imputation and analysis models are correctly specified and are compatible with the analysis model [42]. A limitation of the use of DAGs is that they give no information about the size or magnitude of any bias. Depending on the exact parameterisation of the relationships between variables, there will be occasions when the algorithm suggests that MI estimates in the full sample may be biased, and yet in practice the bias might be negligible. This may arise from specific patterns of missing data within a given dataset, for example in Figure 1A if no participant had both $X$ and $Y$ missing then MI of the whole dataset would unbiasedly estimate the target parameter, despite the conclusions made by the algorithm. Future work could focus on considerations when selecting the subsample to use for MI – these could include the amount of missingness in each variable, and the fraction of missing information. We have already noted that imputing just the outcome using only analysis model variables will not aid in the precision of the target parameter estimate [2] which suggests that some subsample selections will be more useful than others. We have assumed for convenience that the same auxiliary variables are included in all imputation models, though this may not be necessary. Further work is needed to aid the identification of appropriate and necessary auxiliaries in the context of subsample MI.

# Conclusion

We have provided here an easy-to-implement algorithm to enable researchers to decide how plausible it is that MI applied to the full dataset will estimate a target exposure-outcome association without bias, for a given assumed causal structure. Application of a previously-derived concept of "subsample MI" [24] provides a rationale for exploring subsamples in which MI is valid, even when MI in the whole dataset is not.

# Acknowledgements

Our initial ideas, including the directed acyclic graphs in Figures 1 and 3, were inspired by a blog post by Paul Allison: The Peculiarities of Missing at Random. Available from: https://statisticalhorizons.com/missing-at-random/.

# Funding

PMD, RAH, JH and KT work in the Medical Research Council Integrative Epidemiology Unit at the University of Bristol which is supported by the UK Medical Research Council and the University of Bristol (Grant ref: MC_UU_00032/02). PMD is supported by the NIHR Biomedical Research Centre at the University of Bristol and University Hospitals Bristol and Weston NHS Foundation Trust (Grant ref: NIHR203315). RAH is supported by a Sir Henry Dale Fellowship that is jointly funded by the Wellcome Trust and the Royal Society (Grant ref: 215408/Z/19/Z). MBM was supported by National Institutes of Health grants R01LM013866, UL1TR003142, P30CA124435, and P30DK116074. For the purpose of Open Access, the author has applied a CC BY public copyright licence to any Author Accepted Manuscript version arising from this submission.

# Figures

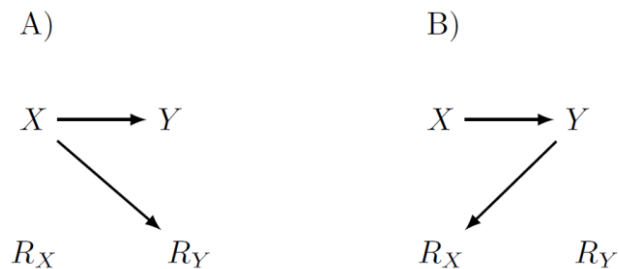

Figure 1: Directed acyclic graphs (DAGs) of two examples. The variable X is the exposure, Y is the outcome and $R_X$ and $R_Y$ represent response indicator variables equal to 1 when X and Y are observed (i.e., they are not missing) respectively. We do not include any boxes around variables to allow the DAGs to represent the data generating mechanism and not a specific estimator (such as complete records analysis or multiple imputation). For A, the target parameter (the regression coefficient for the effect of X on Y) will be estimated without bias using complete records analysis, while in B this target parameter will be estimated with bias as missingness is dependent on the outcome. In both A and B, the target parameter may be estimated with bias when using multiple imputation including all study participants.

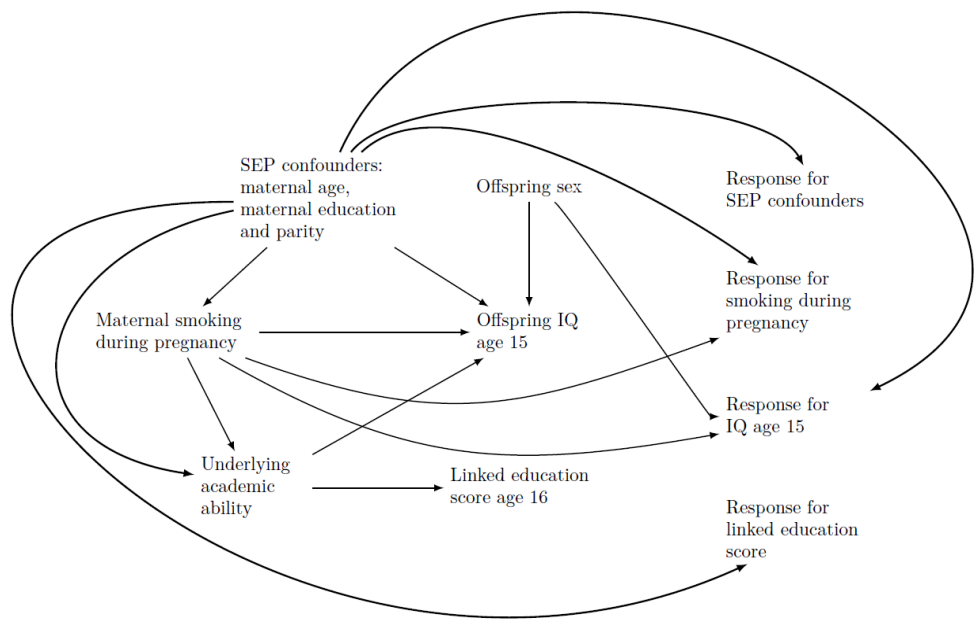

Figure 2: Directed acyclic graph of the assumed relationships between variables and missing data indicators in an applied example from the Avon Longitudinal Study of Parents and Children investigating the effect of maternal smoking during pregnancy (exposure) on offspring intelligence quotient (IQ) scores at age 15 (outcome). In this example the analysis model is a linear regression of the outcome on the exposure adjusted for SEP confounders and offspring sex. Linked education score at age 16 is being used as an auxiliary variable for the incomplete outcome. Underlying academic ability is an unmeasured variable.

## A) Scenario 1

**Steps D1-D3**

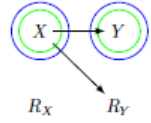

$\Phi = \{X, Y\}$.

**Steps D4-D7**

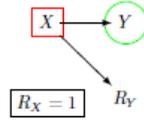

Condition on complete $X$, impute $Y$ (as shown) or condition on complete $Y$ and impute $X$.

## B) Scenario 2

**Steps D1-D3**

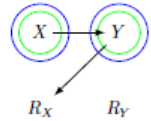

$\Phi = \{X, Y\}$.

**Steps D4-D7**

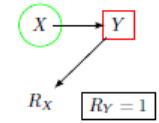

Condition on complete $Y$, impute $X$.

## Algorithm implementation for scenario 1

**Step D0.** Draw a DAG of the assumed causal structure. Identify complete variables with a red box and incomplete variables with a green outline. There are no complete variables in this example.

**Step D1.** Identify all variables with an open path to a response indicator, conditional on all complete variables (of which there are none here). This is set $\Phi$. Here $\Phi$ consists of variables $X$ and $Y$, shown with a blue outline.

**Step D2.** Remove from set $\Phi$ all complete variables (i.e. leave only those variables with a blue outline and not a red box) – $X$ and $Y$ are incomplete so $\Phi$ contains variables $X$ and $Y$.

**Step D3.** MI will therefore not be valid using all available data to impute $Y$ and $X$ (because $\Phi$ is not the empty set).

**Step D4.** Remove all blue and green outlines from all variables in the DAG. Choose to subsample on complete $X$, i.e. $Q = \{X\}$ and $P = \{Y\}$. Add a red box to $X$ (as it is complete in this subsample), and a green outline to $Y$ (as it is incomplete). We draw a black box around $R_X$ and indicate that it takes the value 1 for this subsample.

**Step D5.** Identify any open paths between $Y$ and $R_Y$, conditional on $X$.

**Step D6.** There are no such open paths, so MI of $Y$ in the subsample in which $X$ is complete will unbiasedly estimate the target parameter.

**Alternative Steps D4-D7.** Choose $Q = \{Y\}$ and $P = \{X\}$. Add a red box to $Y$, a green outline to $X$, and a black box around $R_Y$ and indicate that it takes the value 1 for this subsample. There are no open paths between $X$ and $R_X$ conditional on $Y$, so MI of $X$ in the subsample in which $Y$ is complete will unbiasedly estimate the target parameter

## Algorithm implementation for scenario 2

**Step D0-D3.** $\Phi$ consists of the variables $X$ and $Y$. MI applied to the whole dataset may estimate the target parameter with bias as $\Phi$ is not the empty set.

**Step D4-D7.** Remove all blue and green outlines from the DAG. Add a red box to $Y$ and a black rectangle around $R_Y$ to indicate we have subsampled on $Y$ being complete. Add a green outline around $X$ to indicate it is still incomplete. We identify there are no paths from $X$ to $R_X$ conditional on $Y$, so MI of $X$ in the subsample in which $Y$ is complete will unbiasedly estimate the target parameter. It is not possible to use subsampling on complete $X$ in this scenario because $R_X$ is dependent on the outcome $Y$.

*Figure 3: Examples of the algorithm's implementation being applied to the directed acyclic graphs presented in Figure 1A and B. The target parameter is the regression coefficient for X in a regression of Y on X.*

## Algorithm implementation

**Scenario 3**

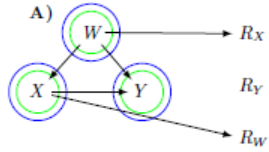

A)

$\Phi = \{W, X, Y\}$.

**Step D0.** Draw the DAG, outlining all complete variables in red (none) and all incomplete variables in green ($W$, $X$ and $Y$).

**Step D1.** Draw a blue outline around all variables with an open path to any response indicator, conditional on variables with a red box (in this case $W$ and $X$ and $Y$ all have blue outlines).

**Step D2 and D3.** Remove all complete variables from $\Phi$ (no variables removed here). $\Phi$ is the set of variables with a blue and green outline ($W$, $X$ and $Y$). MI using all available data is therefore not valid as $\Phi$ is not the empty set.

Steps for identifying a subsample of observed values in which MI can be used to estimate the target parameter without bias (Plot B and C)

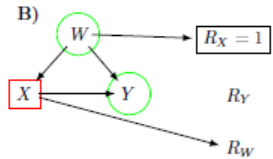

B)

$P = \{W,Y\}, Q = \{X\}$. Condition on complete $X$, impute $Y$ and $W$.

**Step D4.** Remove all blue and green outlines from variables. Select one variable to make complete (i.e. for plot B, $Q = X, P = \{W,Y\}$ and for plot C, $Q = W, P = \{X,Y\}$). Add a red box to the 'complete' variable and draw a box around its response indicator. Add green outlines to the incomplete variables.

**Step D5.** Identify any open paths between variables in $P$ and response indicators for variables in $P$, conditional on variables with a red box.

**Step D6.** There are no open paths from $Y$ or $W$ to $R_Y$ or $R_W$ in plot B or from $X$ or $Y$ to $R_X$ or $R_Y$ in plot C. The target parameter can be estimated without bias using MI of $Y$ and $W$ in the subsample with complete $X$ in plot B and using MI of $X$ and $Y$ in the subsample with complete $W$ in plot C.

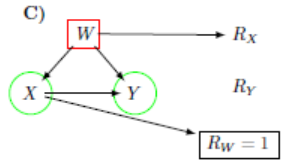

C)

$P = \{X,Y\}, Q = \{W\}$. Condition on complete $W$, impute $Y$ and $X$.

*Figure 4: Worked example of algorithm implementation involving three variables where W is a confounder of X and Y, with W causing missingness in X, X causing missingness in W, and missingness in Y not caused by any variable. The target parameter is the regression coefficient for X in a regression of Y on X and W.*

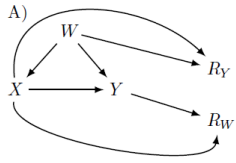 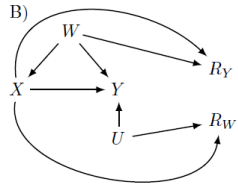 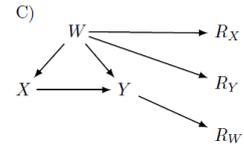

A) Condition on complete $Y$, impute $W$.  B) No possible subsampling sets exist.  C) Condition on complete $Y$ and $X$, impute $W$.

Figure 5: Directed acyclic graphs for the further examples involving three analysis model variables. In each the target parameter is the regression coefficient for X in a regression of Y on X and W.

# Supplementary materials

Using directed acyclic graphs to determine whether multiple imputation or subsample multiple imputation estimates of an exposure-outcome association are unbiased

**Contents**

# Extended background

Consider two scenarios, presented in Figure 1Figure 1, that were inspired by an example provided in an online blog by Paul Allison [43] and have been explored by Mohan and Pearl previously (see Figure 4 of their paper) [8]. In both scenarios there are two variables, an exposure variable $X$ that causes an outcome variable $Y$, both of which are partially observed with response indicators $R_X$ and $R_Y$ equal to 1 when that variable is observed. Note that we have drawn Figure 1 to represent the data generating mechanism only and so no boxes, representing conditioning, are included. To use the DAG to explore bias in a CRA, in accordance with DAG convention, boxes would be drawn around variables which are conditioned on in the analysis – in this case, only the exposure variable $X$. In the presence of incomplete data, a CRA also conditions on the response indicators, by restricting the analysis to the sample in which $X$ and $Y$ are observed (i.e. $R_X$ and $R_Y$ are both equal to 1). In the first scenario (Figure 1A) missingness in $Y$ is caused by $X$ (indicated by the arrow from $X$ to $R_Y$), while in the second (Figure 1B) missingness in $X$ is caused by $Y$. Based on the DAG it is easy to establish that the regression coefficient for the exposure-outcome effect will be unbiased using CRA for the first scenario as missingness does not depend on the outcome (Figure 1A). In the second scenario (Figure 1B) the estimate for the exposure-outcome effect may be biased using CRA, because missingness in $X$ is dependent on the outcome variable. In both scenarios the data are not MAR (or z-MAR as the m-backdoor criterion is not met), therefore the estimate of the exposure-outcome coefficient may be biased using MI. This may be easy to see for these simple examples, as the missingness indicators are dependent on incomplete variables. However, in complex scenarios, with more incomplete analysis model variables, visually identifying such dependencies becomes more difficult.

# Non-graphical version of the algorithm

We provide below a non-graphical version of the algorithm that does not rely on DAGs for its implementation. However, it is important to note here that 1) we are still making assumptions about the data generating mechanism and not looking at associations in the actual data (i.e. we are implicitly relying on a DAG specifying the assumed causal structure, even if one has not been drawn), and 2) the user still needs to consider unmeasured variables $U$ in the process.

> A1) Identify a subset $\Phi$ which contains all variables that are associated with the response indicator for any variable, conditional on all complete variables – i.e. $\Phi$ includes all variables that are dependent with some $R_J$, where $J \in Z'$, conditional on $Z$.
>
> A2) Remove all complete and all unmeasured variables from $\Phi$. $\Phi$ is now the subset of incomplete measured variables that are dependent with the response indicator for any incomplete variable, i.e. the set of MNAR-inducing variables.
>
> A3) If $\Phi$ is now empty, then the MI estimate will be unbiased using the entire dataset (i.e. imputing all incomplete variables in $Z'$). This is because no variables are in $\Psi$ and so the data are z-MAR [23].

If $\Phi$ is not the empty set, we now wish to identify whether subsample MI, i.e. partitioning the incomplete variables Z' into subsets $P$ and $Q$, followed by MI of the variables in $P$ conducted in the subsample in which $Q$ is complete (i.e. individuals $i$ for whom $R_{Ji} = 1$ for all $J \in Q$), would unbiasedly estimate the target parameter. Using the criteria from Little and Zhang, the subsample MI estimate is unbiased if both (1) Y is not related to any response indicators for variables in $Q$, conditional on the analysis model covariates $X$ and $W$, and (2) all variables in $P$ are z-MAR in this subsample [24]. For any given choice of $Q$, our algorithm proceeds as follows:

> A4) Restrict to the subsample of individuals with observed values for all variables contained in $Q$.
>
> A5) In this restricted sample, apply steps A1-A3. This is to check whether $P$ are subsample-MAR (i.e. z-MAR in the subsample of individuals with complete data for all variables in $Q$).
>
> A6) If $\Phi$ is now empty, then $P$ are subsample-MAR, and as $Y$ is unrelated to membership of this subsample, then MI in this subsample is valid and will unbiasedly estimate the target parameter.

A7) If $\Phi$ is not empty, $P$ are not subsample-MAR so MI in this subsample is not valid. Alternative approaches could include considering different subsamples or identifying further auxiliary variables to block paths between incomplete variables and response indicators.

# Extended further examples

We provide brief discussion for four further examples, three of which are presented in Figure 5 of the main text, and one presented in Supplementary Figure S1 below.

In Figure 5A the confounder $W$ and exposure $X$ cause $R_Y$, both $Y$ and $X$ cause $R_W$, and $X$ is complete. Following steps D1-3, the set $\Phi$ consists of $W, Y$. As this is not the empty set MI applied to the whole dataset is not valid. The set of incomplete variables, $Z'$, also consists of $W, Y$. Separating this set into P, incomplete variables whose response indicator is dependent on Y conditional on analysis model variables, and Q, incomplete variables whose response indicator is independent of Y conditional on analysis model variables, gives $P = W$ and $Q = Y$. We restrict to observed values for all variables in $Q$ (i.e. restrict to observations with observed $Y$), and check whether the variables in $P$ (i.e. $W$) are independent of all response indicators conditional on all complete variables ($Z = X$) and all variables in $Q$ (i.e. $Y$). Restriction to observed $Y$ eliminates the path from $W$ to $R_Y$ in the subsample and there is no open path from W to $R_W$ conditional on $X$ and $Y$ (which would be included in the imputation model for $W$), meaning that this check is passed. The data are therefore z-MAR in the subsample and $W$ can be imputed in the subsample with complete $Y$. This provides an example where restricting to a complete outcome variable will result in unbiased estimation of the target parameter via subsample MI.

Figure 5B is the same as Figure 5A, except instead of a direct path from $Y$ to $R_W$, there is an unmeasured common cause $U$ of $Y$ and $R_W$. As before $W$ and $X$ cause $R_Y$, $X$ also causes $R_W$, $X$ is complete, and the set $\Phi$ initially consists of $W, Y, U$ before removal of U (unmeasured) to leave $\Phi$ containing $W, Y$ so MI in the whole sample is not valid. Separating the incomplete variables $Z'$ into $P$ (response dependent on $Y$) and $Q$ (response independent of $Y$) again yields $P = W$ and $Q = Y$, i.e. imputing W in the subsample with complete values of $Y$. However, the check for independence between $W$ and all response indicators fails as $Y$ is a collider for $W$ and the unmeasured variable $U$, resulting in an open path between $W$ and $R_W$ when we condition on $Y$ (which is now a complete variable in this subsample). The data are therefore not z-MAR in the subsample with complete $Y$ and it is not possible to unbiasedly estimate the target parameter using subsample MI using any subsample of observed variables. The comparison between Figure 5A and B highlights the importance of temporality of variables. It is more likely that there is common cause of $Y$ and response for the earlier occurring variable $W$ than for $Y$ to directly cause response in $W$. It is therefore possible to subsample on observed values of the outcome but is unlikely to be feasible in practice due to collider bias. If instead the variable $U$ is measured and complete, then it can be included in the imputation model as an auxiliary variable to close the path between $W$ and $R_W$ conditional on $Y$, making the data z-MAR in the subsample with complete $Y$.

In Figure 5C the variable $X$ is additionally incomplete. $W$ causes $R_X$ and $R_Y$, and Y causes $R_W$. Following steps D1-3, the set $\Phi$ consists of $X, W, Y$ and MI in the full sample is not valid. We note that we cannot subsample on complete $W$, because $Y$ causes $R_W$. Instead, we could subsample on complete values for the variables $X$ and $Y$ - partitioning incomplete variables $Z'$ into $P = \{W\}$ and $Q = \{Y, X\}$. Subsampling on complete $Y$ and $X$ (i.e. restricting to those individuals with both $Y$ and $X$ fully observed), results in the data being z-MAR in the subsample as $W$ is no longer related to a response indicator for an incomplete variable, conditional on $X$ and $Y$ being complete in this subsample. This scenario shows that it is possible, and sometimes necessary, to subsample on observed values of multiple incomplete variables to unbiasedly estimate the target parameter via subsample MI, though we note that the temporality issue described above in Figure 5B is still relevant here.

Finally, Supplementary Figure S1 is an example provided in our previous work exploring incomplete auxiliary variables [25]. In this example $X$ causes $Y$ and $R_Y$, and an auxiliary variable $A$ causes $Y$ and its own missingness ($R_A$). In this example the target parameter is the unconditional regression coefficient of $Y$ on $X$. In our previous work we showed via simulation that the target parameter is unbiasedly estimated using CRA and MI excluding the auxiliary but not using MI including the auxiliary. Implementing the algorithm in the absence of $A$ gives $\Phi = \emptyset$ and hence MI excluding the auxiliary in the whole dataset is valid. Implementing the algorithm in the presence of the auxiliary gives $\Phi = \{Y, A\}$ and hence MI including the auxiliary in the whole dataset is not valid. Our algorithm steps D0-3 predict the previous simulation finding for both settings. We then investigate separating $Z' = \{Y, A\}$ into $P = \{A\}$ and $Q = \{Y\}$. Restricting to complete $Q$ (i.e. complete $Y$) simply gives the CRA sample and so subsample MI in this example will not provide any efficiency gains beyond CRA. Separating $Z' = \{Y, A\}$ into $P = \{Y\}$ and $Q = \{A\}$ and restricting to complete $Q$ (i.e. complete A) may estimate the target parameter with bias because $R_A$ is dependent on $Y$, conditional on $X$ (we do not condition on $A$ here because it is not in the analysis model). Thus, in this example we would choose to use CRA to estimate the target parameter as 1) MI of the outcome in the whole dataset using no auxiliaries would also be unbiased but less precise than CRA [2], 2) MI of the outcome in the whole dataset including the auxiliary would be biased and 3) subsample MI would either be biased ($Q = \{A\}$) or less precise than CRA ($Q = \{Y\}$).

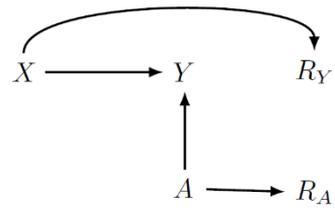

*Figure S1: Additional example including an incomplete auxiliary variable.*

# Figures showing application of algorithm to the motivating example

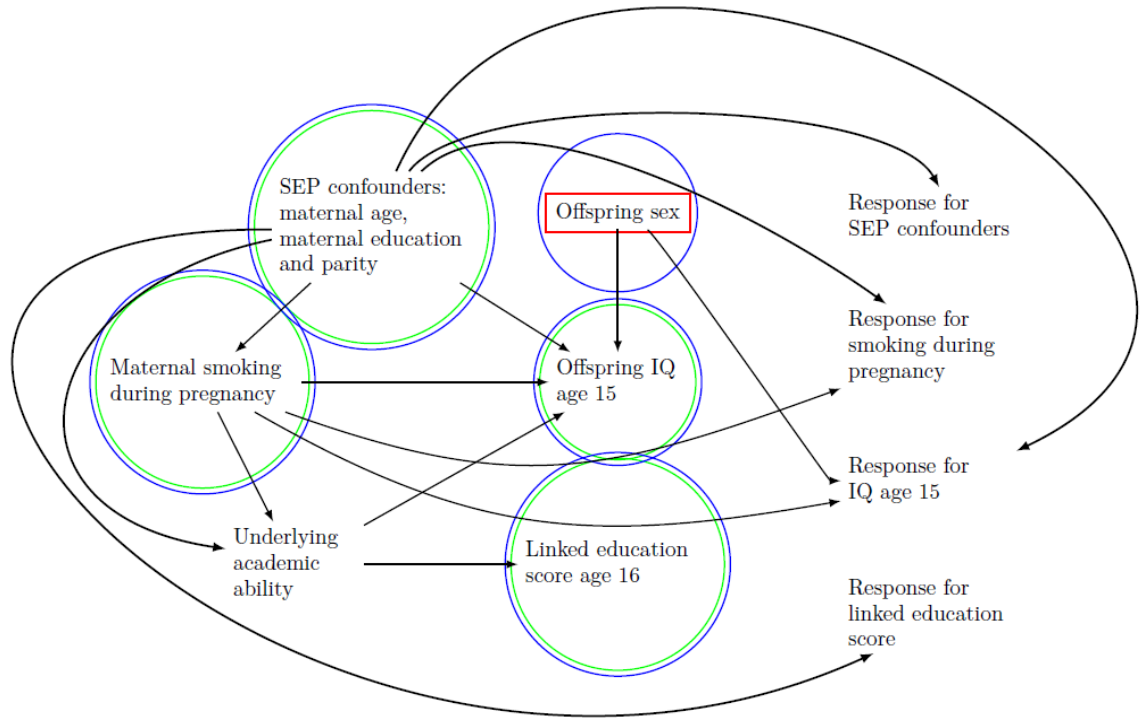

$\Phi = \{$maternal smoking during pregnancy, IQ at age 15, SEP confounders, linked education score$\}$

*Figure S2: Application of steps D0-D3 of the algorithm to the motivating example.*

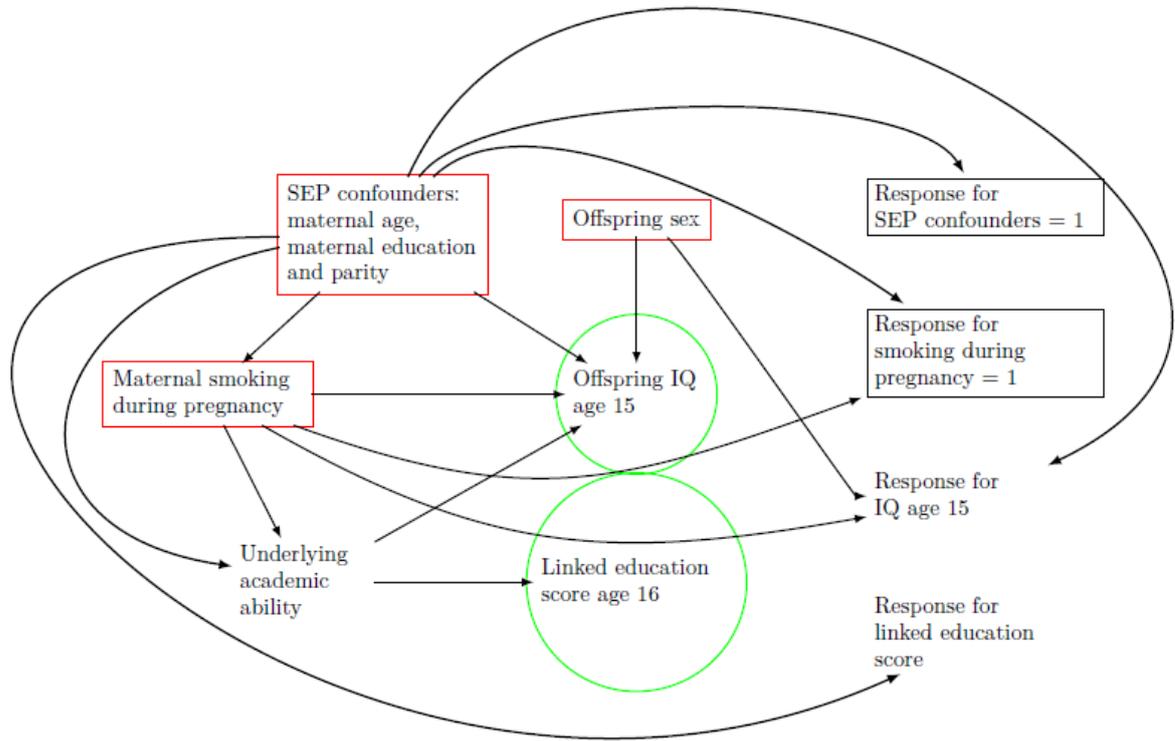

$P = \{\text{IQ at age 15, linked education score}\}$

$Q = \{\text{maternal smoking during pregnancy, SEP confounders}\}$

*Figure S3: Application of steps D4-D7 of the algorithm to the motivating example.*

# Algorithm applied to canonical missingness DAGs of Moreno-Betancur et al 2018 [5]

- Analysis is Y | X, Z1, Z2, i.e. the target parameter is the regression coefficient for X in the regression of Y on X adjusted for Z1 and Z2.
- Z1 variables are fully observed.
- No auxiliary variables
- U and W are unmeasured

A)

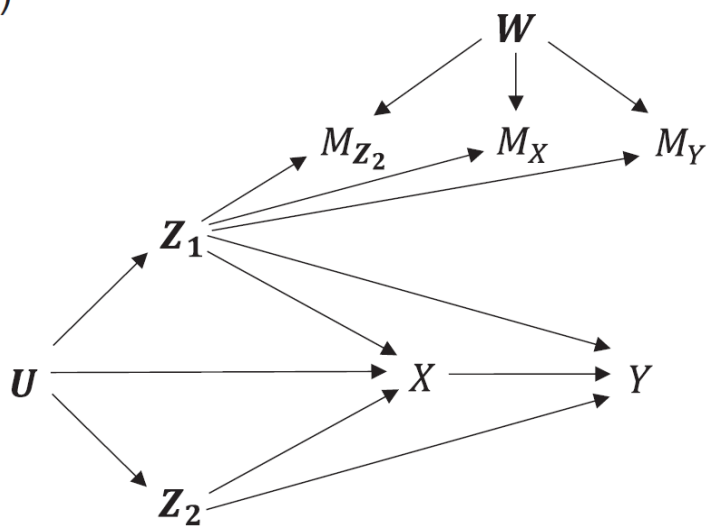

- Missingness in Y depends on Z1.
- Missingness in X depends on Z1.
- Missingness in Z2 depends on Z1.

Apply algorithm to full sample

A1) Φ=Z1, W

A2) Φ ={}

A3) Φ is empty; exit algorithm. Impute Y, X and Z2 based on entire sample.

**Can apply standard MI to entire sample since the data are z-MAR given fully observed Z1.**

B)

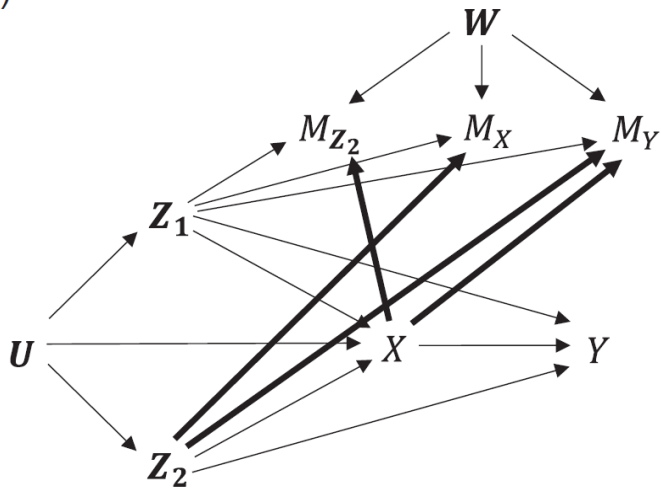

- Missingness in Y depends on X, Z1, Z2.
- Missingness in X depends on Z1, Z2.
- Missingness in Z2 depends on X, Z1.

Apply algorithm to full sample

A1) Φ =X, Z1, Z2, Y, U, W

A2) Φ =X, Z2, Y

A3) Φ is not empty; estimate from MI applied to the full dataset will be biased.

Apply algorithm to subsample $M^X$=0 and $M^{Z_2}$=0

A4) Q =X, $Z_2$, P=Y

A5) Restrict to $M^X$=0 and $M^{Z_2}$=0

A6) Within this subsample, Φ is empty (i.e. all incomplete variables are d-separated from response indicators by complete variables). Estimate from MI of Y within the subsample $M^X$=0 and $M^{Z_2}$=0 will be unbiased.

- **In the full sample, the data are not z-MAR.**
- **Within subsample $M^X = 0, M^{Z_2} = 0$, the data are z-MAR given fully observed variables.**
- **No smaller subsamples can be found where the estimate from MI will be unbiased.**

C)

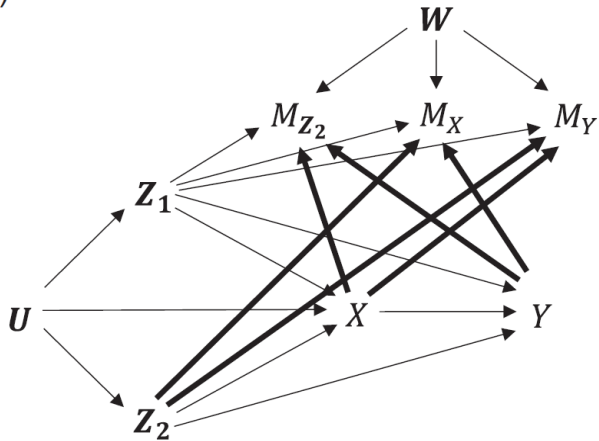

- Missingness in Y depends on X, Z1, Z2.
- Missingness in X depends on Y, Z1, Z2.
- Missingness in Z2 depends on X,Y, Z1.

Apply algorithm to full sample

A1) Φ =X,Y,Z1,Z2,W,U

A2 Φ =X,Y, Z2

A3) Φ is not empty; so the estimate from MI applied to the full dataset may be biased.

Apply algorithm to subsample $M^Y$=0

A4) Q=Y, P =X, $Z_2$,

A5) Restrict to $M^Y$=0

A6) Within this subsample, Φ = { X, $Z_2$}

A7) Φ is not empty , so the estimate from MI of X, $Z_2$ within the subsample with complete Y ( $M^Y$=0) will be biased.

Applying algorithm to other partitions is not possible because Y causes response indicators for X and $Z_2$.

- **In the full sample, the data are not z-MAR.**
- **Within subsample $M^Y = 0$, the data are not z-MAR for missing data pattern jointly missing X and Z2.**

D)

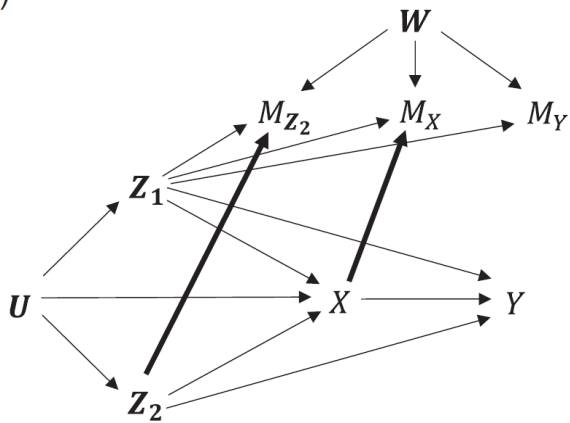

- Missingness in Y depends on Z1.
- Missingness in X depends on X, Z1.
- Missingness in Z2 depends on Z1, Z2.

Apply algorithm to full sample

A1) Φ =X, Z1, Z2, Y, U, W

A2) Φ =X, Z2, Y

A3) Φ is not empty; estimate from MI applied to the full dataset will be biased.

Apply algorithm to subsample $M^X$=0 and $M^{Z_2}$=0

A4) Q =X, $Z_2$, P=Y

A5) Restrict to $M^X$=0 and $M^{Z_2}$=0

A6) Within this subsample, Φ is empty (i.e. all incomplete variables are d-separated from response indicators by complete variables). Estimate from MI of Y within the subsample $M^X$=0 and $M^{Z_2}$=0 will be unbiased.

- **In the full sample, the data are not z-MAR due to paths from incomplete variables X and Z2 to response indicators.**
- **Within subsample $M^X = 0, M^{Z_2} = 0$, the data are z-MAR given fully observed variables.**
- **No smaller subsamples can be found where the estimate from MI will b unbiased.**

E)

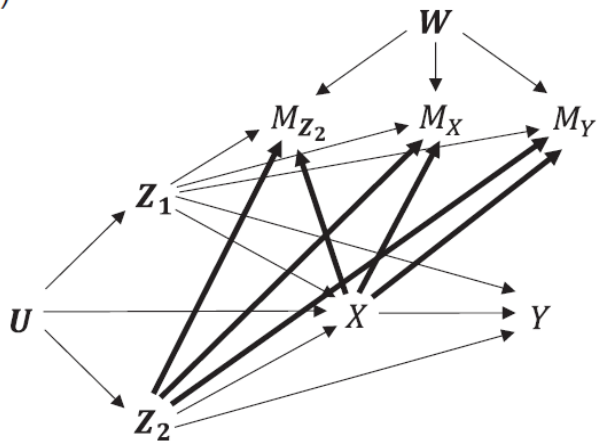

- Missingness in Y depends on Z1, X, and Z2.
- Missingness in X depends on X, Z1, Z2.
- Missingness in Z2 depends on X, Z1, Z2.

Apply algorithm to full sample

A1) Φ =X, Z1, Z2, Y, U, W

A2) Φ =X, Z2, Y

A3) Φ is not empty, so MI applied to the full sample may estimate the target parameter with bias; proceed to step 4.

A3) Φ is not empty; estimate from MI applied to the full dataset will be biased.

Apply algorithm to subsample $M^X$=0 and $M^{Z_2}$=0

A4) Q =X, $Z_2$, P=Y

A5) Restrict to $M^X$=0 and $M^{Z_2}$=0

A6) Within this subsample, Φ is empty (i.e. all incomplete variables are d-separated from response indicators by complete variables). Estimate from MI of Y within the subsample $M^X$=0 and $M^{Z_2}$=0 will be unbiased.

- **In the full sample, the data are not z-MAR due to paths from incomplete variables X and Z2 to response indicators. Within subsample $M^X = 0, M^{Z_2} = 0$, the data are z-MAR given fully observed variables.**

F)

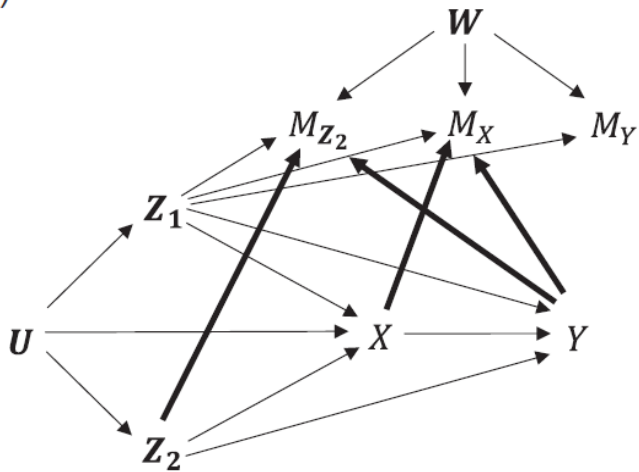

- Missingness in Y depends on Z1.
- Missingness in X depends on Y, X, Z1.
- Missingness in Z2 depends on Y, Z1, Z2.

Apply algorithm to full sample

A1) Φ =Y, X, Z1, Z2, U, W

A2) Φ =Y, X, Z2

A3) Φ is not empty; so, the estimate from MI applied to the full dataset may be biased.

Apply algorithm to subsample $M^Y$=0

A4) Q=Y, P =X, Z$_2$,

A5) Restrict to $M^Y$=0

A6) Within this subsample, Φ = { X, Z$_{2,}$}

A7) Φ is not empty, so the estimate from MI of X, Z$_2$ within the subsample with complete Y ( $M^Y$=0) will be biased.

Applying the algorithm to other partitions is not possible because Y causes response indicators for X and Z$_2$.

- **In the full sample, the data are not z-MAR due to paths from incomplete variables X, Z2 and Y to response indicators. Within subsample $M^Y = 0$, the data the data are not z-MAR due to paths from incomplete variables X and Z2 to response indicators.**

G)

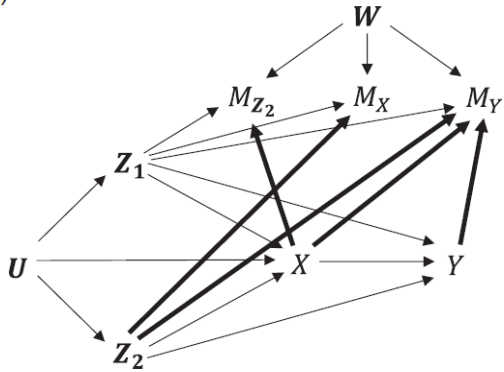

- Missingness in Y depends on Y, X, Z1, Z2.
- Missingness in X depends on Z1, Z2.
- Missingness in Z2 depends on X, Z1.

Missingness in Y depends on itself. There is no need to apply the full algorithm, as we know that Y cannot be imputed (as incomplete Y will always cause the data to not be z-MAR) and we cannot apply MI in the sample in which Y is complete (because Y causes inclusion in that subsample). Thus, MI applied to the full sample, or to any subsample, may estimate the target parameter with bias.

H)

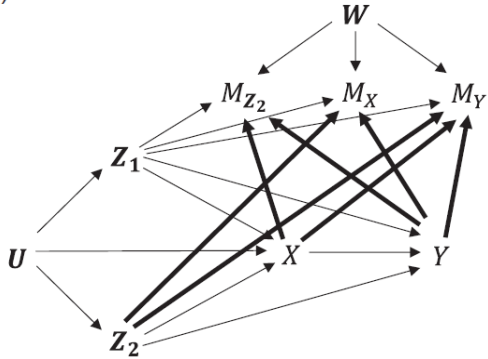

- Missingness in Y depends on Y, X, Z1, Z2.
- Missingness in X depends on Y, Z1, Z2.
- Missingness in Z2 depends on Y, X, Z1.

Missingness in Y depends on itself. There is no need to apply the full algorithm, as we know that Y cannot be imputed (as incomplete Y will always cause the data to not be z-MAR) and we cannot apply MI in the sample in which Y is complete (because Y causes inclusion in that subsample). Thus, MI applied to the full sample, or to any subsample, may estimate the target parameter with bias.

| 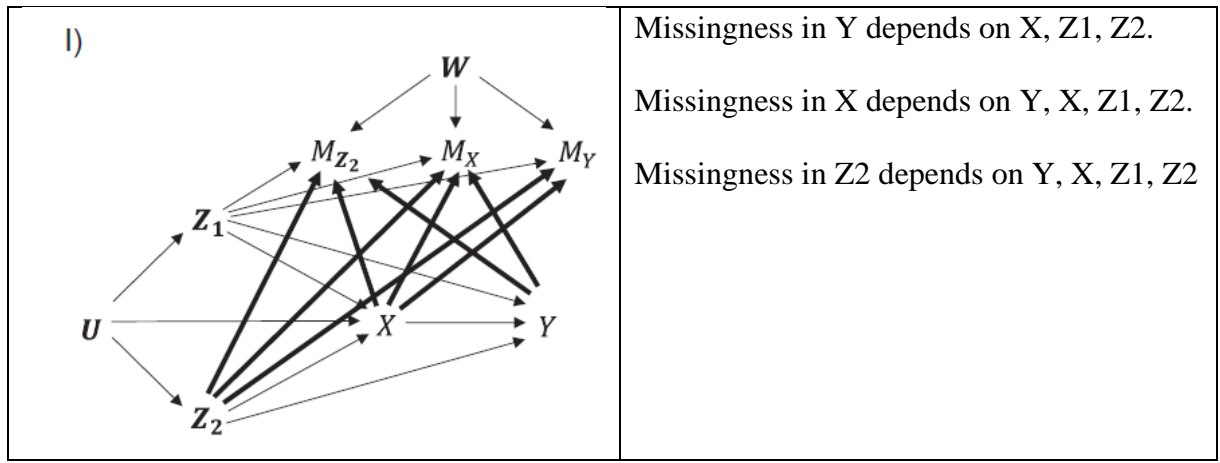 | Missingness in Y depends on X, Z1, Z2.<br><br>Missingness in X depends on Y, X, Z1, Z2.<br><br>Missingness in Z2 depends on Y, X, Z1, Z2 |

Apply algorithm to full sample

A1) Φ =Y, X, Z1, Z2, U, W

A2) Φ =Y, X, Z2

A3) Φ is not empty; so, the estimate from MI applied to the full dataset may be biased.

Apply algorithm to subsample $M^Y$=0

A4) Q=Y, P =X, Z₂,

A5) Restrict to $M^Y$=0

A6) Within this subsample, Φ = { X, Z₂}

A7) Φ is not empty, so the estimate from MI of X, Z₂ within the subsample with complete Y ( $M^Y$=0) will be biased.

Applying algorithm to other partitions is not possible because Y causes response indicators for X and Z₂.

| 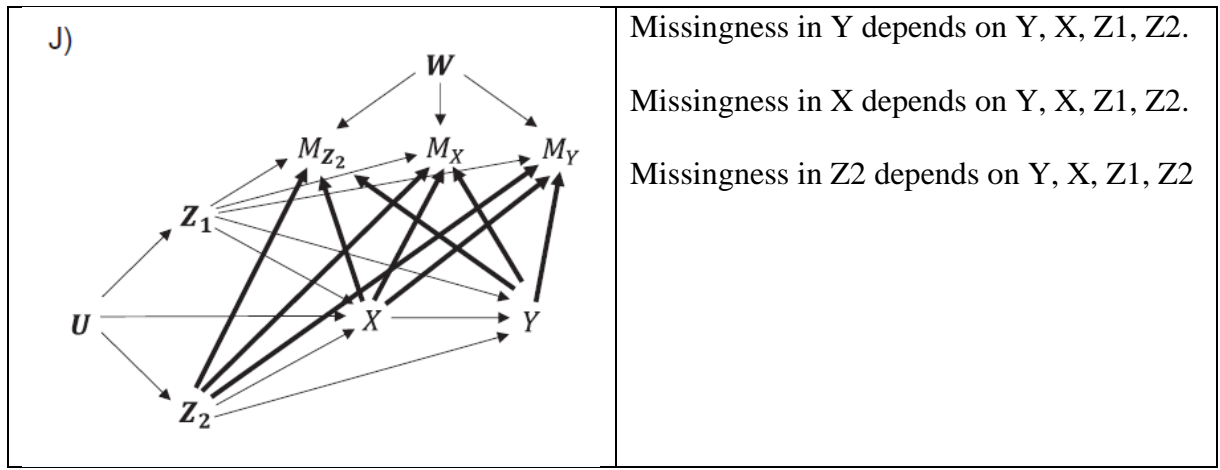 | Missingness in Y depends on Y, X, Z1, Z2. Missingness in X depends on Y, X, Z1, Z2. Missingness in Z2 depends on Y, X, Z1, Z2 |

Missingness in Y depends on itself. There is no need to apply the full algorithm, as we know that Y cannot be imputed (as incomplete Y will always cause the data to not be z-MAR) and we cannot apply MI in the sample in which Y is complete (because Y causes inclusion in that subsample). Thus, MI applied to the full sample, or to any subsample, may estimate the target parameter with bias.

# Simulation study

We describe below the methods and results for a simulation study (using the ADEMP framework [44]) exploring the scenarios described in figures 1/3, 4 and 5. All simulations were conducted in Stata 18.5MP. Code for the simulations can be found at https://github.com/pmadleydowd/Subsample-MI.

## Simulation study methods

### Aims

The simulation study aims to show the bias associated with complete records analysis, MI using all available data, and subsample MI for the scenarios described in Figures 1/3, 4 and 5 in the main text.

### Data-generating mechanisms

Throughout the simulation studies, the target parameter of interest (regression coefficient for $Y$ on $X$, adjusted for $W$ where appropriate) has a true value of 0.15. The variables $U$, $W$, $X$ and $Y$ were simulated to have mean 0 and variance 0.5.

**Figure 1A/B, 3A/B**

We then simulated $X$, $Y$, probability of observed $X$ and probability of observed $Y$ as:

$X \sim N(1, 0.5)$

$Y = 0.15X + 0.85 + (\sqrt{0.5^2 - 0.15^2}) \times \varepsilon$ where $\varepsilon \sim N(0,1)$

$P(\text{observe } X) = 0.9$ if $Y < \text{median}(Y)$, $0.1$ otherwise

$P(\text{observe } Y) = 0.5$

**Figure 4**

We simulated $W$, $X$, $Y$, and the probability of observing each of $W$, $X$ and $Y$ as:

$W \sim N(1, 0.5)$

$X = \sqrt{0.5}W + (1 - \sqrt{0.5}) + (\sqrt{0.5^2}) \times 0.5 \times \varepsilon$ where $\varepsilon \sim N(0,1)$

$Y = 0.15X - 0.5W + 1.35 + \left(\sqrt{0.25 - \left(0.15^2 + \left(\frac{0.5^2}{4}\right) - 2 \times 0.15 \times 0.5 \times \sqrt{0.5}/4\right)}\right) \times \varepsilon$ where $\varepsilon \sim N(0,1)$

$P(\text{observe } X) = 0.9$ if $W < \text{median}(W)$, $0.1$ otherwise

$P(\text{observe } W) = 0.9$ if $X < \text{median}(X)$, $0.1$ otherwise

$P(\text{observe } Y) = 0.5$

**Figure 5A**

We simulated $W, X, Y$ as for scenario 3, and the probability of observing $W, X$ and $Y$ as:

$P(\text{observe } W) = 0.9$ if $(X < \text{median}(X)) \& (Y < \text{median}(Y))$, $0.1$ if $(X > \text{median}(X)) \& (Y > \text{median}(Y))$, $0.5$ otherwise

$P(\text{observe } X) = 1$

$P(\text{observe } Y) = 0.9$ if $(X < \text{median}(X)) \& (W < \text{median}(W))$, $0.1$ if $(X > \text{median}(X)) \& (W > \text{median}(W))$, $0.5$ otherwise

**Figure 5B**

We simulated $U, W, X, Y$, and the probability of observing each of $W, X$ and $Y$ as:

$W \sim N(1, 0.5)$

$U \sim N(1, 0.5)$

$X = \sqrt{0.5}W + (1 - \sqrt{0.5}) + (\sqrt{0.5}) \times 0.5 \times$ where $\varepsilon \sim N(0,1)$

$Y = 0.15X - 0.5W + 2 - U \times \left(\sqrt{2\left(0.25 - \left(\frac{0.15^2}{4} + \left(\frac{0.5^2}{4}\right) - 2 \times 0.15 \times 0.5 \times \frac{\sqrt{0.5}}{4}\right)\right)}\right) +$

$\left(0.5 \times \sqrt{2\left(0.25 - \left(\frac{0.15^2}{4} + \left(\frac{0.5^2}{4}\right) - 2 \times 0.15 \times 0.5 \times \sqrt{0.5}/4\right)\right)}\right) \times \varepsilon$ where $\varepsilon \sim N(0,1)$

$P(\text{observe } X) = 1$

$P(\text{observe } W) = 0.9$ if $(X < \text{median}(X))\&(U < \text{median}(U)), 0.1$ if $(X > \text{median}(X))\&(U > \text{median}(U)), 0.5$ otherwise

$P(\text{observe } Y) = 0.9$ if $(X < \text{median}(X))\&(W < \text{median}(W)), 0.1$ if $(X > \text{median}(X))\&(W > \text{median}(W)), 0.5$ otherwise

**Figure 5C**

We simulated $W, X, Y$, as for scenario 3, and the probability of observing $W, X$ and $Y$ as:

$P(\text{observe } X) = 0.8$ if $(W > 70\text{th centile}(W), 0.4$ otherwise

$P(\text{observe } W) = 0.9$ if $(Y < \text{median}(Y)), 0.1$ otherwise

$P(\text{observe } Y) = 0.9$ if $(W < \text{median}(W), 0.1$ otherwise

## Estimand/target of analysis

In each case, the estimand of interest was the coefficient for the linear regression of $Y$ on $X$ (unconditional for Figure 1/3 and conditional on $W$ for Figures 4 and 5).

## Methods to be evaluated

In each case, the complete records analysis and the MI using all available data were performed, along with the specified subsample-MI analyses.

## Performance measures

500 simulations were carried out for each example. We estimate 1) the average bias across simulations of the coefficient for the linear regression of Y on X (conditional on W where relevant) relative to the true value of 0.15, and 2) the empirical standard error of the bias across simulations.

# Simulation study results

*Table S1: Results from Simulations for one set of parameters for Figures 1A-5C, in all cases true parameter value is 0.15.*

| Scenario | Target parameter biased or unbiased using specified method as predicted by DAG/algorithm | | | Average bias in coefficient for regression of Y on X given W across 500 simulations (empirical SE of bias across 500 simulations) | | | |
|---|---|---|---|---|---|---|---|
| | CRA | MI in whole sample | Subsample MI | Bias in CRA | Bias in MI in whole sample | Subsample MI "Complete" variable(s) | Bias |
| Figure 1A and 3A | Unbiased | Biased | Unbiased conditional on complete X or complete Y | 0.005 (0.086) | -0.034 (0.072) | Y | 0.003 (0.085) |
| | | | | | | X | 0.005 (0.087) |
| Figure 1B and 3B | Biased | Biased | Unbiased conditional on complete Y | -0.060 (0.052) | -0.039 (0.073) | Y | -0.002 (0.085) |
| | | | | | | X | -0.059 (0.052) |
| Figure 4A, B and C | Unbiased | Biased | Unbiased conditional on complete X, complete W, or complete X and W | 0.003 (0.118) | 0.073 (0.100) | X | -0.002 (0.090) |
| | | | | | | W | 0.003 (0.119) |
| | | | | | | X and W | 0.003 (0.112) |
| Figure 5A | Biased | Biased | Unbiased conditional on complete Y | -0.077 (0.083) | -0.022 (0.076) | Y | 0.000 (0.069) |
| | | | | | | W | -0.079 (0.084) |
| Figure 5B | Biased | Biased | Biased | 0.045 (0.082) | 0.063 (0.072) | Y | 0.018 (0.073) |
| | | | | | | W | 0.046 (0.084) |
| Figure 5C | Biased | Biased | Unbiased conditional on complete X and Y | -0.047 (0.005) | -0.402 (0.005) | Y | 0.018 (0.005) |
| | | | | | | W | -0.049 (0.005) |
| | | | | | | X and Y | 0.004 (0.006) |

CRA = Complete records analysis; MI = Multiple imputation; SE = Standard error